\documentclass[aps,superscriptaddress,showpacs,floatfix]{revtex4}
\usepackage{graphicx}
\usepackage{dcolumn}

\usepackage{bm}
\usepackage{CJK}
\usepackage{multirow}
\usepackage[colorlinks,
            linkcolor=red,
            anchorcolor=blue,
            citecolor=green
            ]{hyperref}

\begin{document}

\title{Pairing properties of semilocal coordinate\&momentum-space regularized chiral interactions}
\author{P. Yin}
\affiliation{CAS Key Laboratory of High Precision Nuclear Spectroscopy, Institute of Modern Physics, Chinese Academy of
Sciences, Lanzhou 730000, China}
\author{X. L. Shang\footnote{Corresponding author: shangxinle@impcas.ac.cn}}
\affiliation{CAS Key Laboratory of High Precision Nuclear Spectroscopy, Institute of Modern Physics, Chinese Academy of
Sciences, Lanzhou 730000, China}
\affiliation{School of Nuclear Science and Technology, University of Chinese Academy of Sciences, Beijing 100049, China}
\author{J. N. Hu}
\affiliation{School of Physics, Nankai University, Tianjin 300071, China}
\author{J. Y. Fu}
\affiliation{CAS Key Laboratory of High Precision Nuclear Spectroscopy, Institute of Modern Physics, Chinese Academy of
Sciences, Lanzhou 730000, China}
\affiliation{School of Nuclear Science and Technology, University of Chinese Academy of Sciences, Beijing 100049, China}

\author{E. Epelbaum}
\affiliation{Institut f\"{u}r Theoretische Physik II, Ruhr-Universit\"{a}t Bochum, D-44780 Bochum, Germany}

\author{W. Zuo}
\affiliation{CAS Key Laboratory of High Precision Nuclear Spectroscopy, Institute of Modern Physics, Chinese Academy of
Sciences, Lanzhou 730000, China}
\affiliation{School of Nuclear Science and Technology, University of Chinese Academy of Sciences, Beijing 100049, China}

\begin{abstract}
We investigate the pairing properties of state-of-the-art semilocal coordinate-space and semilocal momentum-space regularized chiral interactions. Specifically, we calculate the pairing gaps in $^3SD_1$ channel of symmetric nuclear matter and in $^1S_0$ and $^3PF_2$ channels of pure neutron matter within the BCS approximation using these chiral interactions. We address the regulator and chiral order dependence of the pairing gaps and compare the pairing properties of the chiral interactions with those of the Argonne v18 (Av18) potential. The effects of the tensor force on the pairing gaps in the $^3SD_1$ and $^3PF_2$ channels are illustrated for both the chiral interactions and the Av18 potential. We evaluate the truncation errors of chiral expansions of the pairing gaps with a Bayesian approach. We find that the pairing gaps converge very well at the higher-order chiral expansions in the $^3SD_1$ and $^1S_0$ channels.
\end{abstract}

\pacs{} \maketitle

\section{Introduction}

Nucleon-nucleon ({\it NN}) interaction, serving as the input of the {\it ab-initio} nuclear many body theory, plays a fundamentally significant role in nuclear physics.
Chiral effective field theory (EFT) allows one to derive {\it NN} interactions based on the underlying fundamental quantum chromodynamics (QCD) and provides a straightforward path to generate consistent and systematically improvable many body interactions and exchange currents~\cite{Epelbaum:2008ga}.

In Refs.~\cite{Epelbaum:2014efa,Epelbaum:2014sza}, a set of semilocal coordinate-space (SCS) regularized chiral EFT {\it NN} interactions were developed up through fifth chiral order (N$^4$LO) using a local regulator for the pion-exchange contributions, which allows one to substantially reduce finite-cutoff artifacts. In particular, the long range contributions are regularized in coordinate space via
$V_{\pi}(\vec{r})\longrightarrow V_{\pi,R}(\vec{r})=V_{\pi}(\vec{r})
\left[1-e^{-\frac{r^2}{R^2}}\right]^n$,
where the cutoff $R$ was chosen in the range of $R=0.8,0.9,1.0,1.1,$ and $1.2$ fm.
The exponent $n$ was set $n=6$, but choosing $n=5$ or $n=7$ led to a comparable description of the phase shifts~\cite{Epelbaum:2014efa}. For contact interactions, a nonlocal Gaussian regulator in momentum space was employed with the cutoff $\Lambda$ being related to $R$ via $\Lambda=2/R$. These novel chiral EFT interactions have been successfully applied to {\it ab-initio} calculations of nuclear structure, nuclear reactions, and nuclear matter~\cite{LENPIC:2015qsz,Maris:2016wrd,LENPIC:2018lzt,Yin:2019kqv,Du:2022zds,Yin:2022zii, Hu:2016nkw,Hu:2019zwa}.
However, the numerical implementation of the three-nucleon potentials with the coordinate-space regulator in the Faddeev and Yakubovsky equations appears to be challenging, in particular as chiral order increases.

Therefore, a new generation of semilocal momentum-space (SMS) regularized chiral EFT {\it NN} interactions was developed in Ref.~\cite{Reinert:2017usi}, where both the short-range and long-range contributions to the interaction are regularized in momentum space. Compared with the SCS regularized interactions, the new SMS regularized interactions remove three redundant short-range operators at N$^3$LO and use the most up-to-date values of the pion-nucleon low-energy constants (LECs) from the Roy-Steiner equation analysis of Refs.~\cite{Hoferichter:2015tha,Hoferichter:2015hva}. Another feature of the SMS regularized interactions is that the highest chiral order, referred to as N$^4$LO+, includes four sixth-order contact interactions in F-waves in order to precisely describe the neutron-proton F-wave phase shifts, which are still not converged at N$^4$LO.
These SMS regularized chiral interactions have also been successfully applied to {\it ab initio} calculations of the nuclear structure and reactions~\cite{Epelbaum:2019zqc,Volkotrub:2020lsr,Urbanevych:2020sjs,Filin:2019eoe,Filin:2020tcs,Maris:2020qne,LENPIC:2022cyu}.

Pairing between nucleons in nuclear matter is key to understand various  phenomena in compact star physics, such as the cooling of new born stars~\cite{Lattimer:1994glx}, the afterburst relaxation in X-ray transients~\cite{Page:2012glx}, and the glitches~\cite{Piekarewicz:2014lba,Shang:2021qhe}. The reliable knowledge of the pairing correlations requires accurate bare {\it NN} interactions as inputs. However, the pairing gaps in nuclear matter have not been well constrained~\cite{Lombardo:2005sw}. In addition, the pairing correlations in the coupled channels, such as $^3SD_1$ and $^3PF_2$, may shed light on the properties of the tensor force. We therefore study in this work the pairing properties of the above addressed chiral EFT interactions in nuclear matter within the BCS approximation. Especially, we focus on the pairing properties in the $^1S_0$, $^3SD_1$, and $^3PF_2$ channels, which are found to be dominant from low to intermediate densities~\cite{Lombardo:2005sw}. We use free single particle spectrum where the only uncertainty of pairing gaps stems from the {\it NN} interactions adopted. Therefore these investigations may reveal essentially the properties of {\it NN} interactions themselves. We defer to use more realistic while sophisticated single particle spectrum in the future, where the effective mass, depletion of the Fermi surface due to short-range correlation effects, and the medium polarization will be taken into account with the Brueckner G-matrix~\cite{Dong:2013sqa,Shang:2013wza,Shang:2013cma,Guo:2018lna}.

\section{Theory and discussion}
Within the BCS approximation, the pairing gap is determined by the following gap equation:
\begin{equation}
\left(
  \begin{array}{c}
    \Delta_L(k) \\
    \Delta_{L+2}(k) \\
  \end{array}
\right)
=-\frac{1}{\pi}\int dk' k'^2
\left(
  \begin{array}{cc}
    V_{L,L}(k,k') & V_{L,L+2}(k,k') \\
    V_{L+2,L}(k,k') & V_{L+2,L+2}(k,k') \\
  \end{array}
\right)
\frac{1}{\sqrt{\xi_{k'}^2+D^2(k')}}
\left(
  \begin{array}{c}
    \Delta_L(k') \\
    \Delta_{L+2}(k') \\
  \end{array}
\right),
\label{eq:gap}
\end{equation}
with
\begin{eqnarray}
  D^2(k) &=& \Delta_L^2(k)+\Delta_{L+2}^2(k), \\
 \xi_{k}&=&\frac{1}{2}(\varepsilon_k^1+\varepsilon_k^2),
\label{eq:Dk}
\end{eqnarray}
 where $\varepsilon_k^1$ and $\varepsilon_k^2$ correspond to the single-particle energies of the two pairing nucleons. The off-diagonal $V_{L,L'}$ vanishes for single channel calculations and the gap equation reduces to $1\times 1$ dimension.

\begin{figure}[tbh]
\begin{center}
\includegraphics[width=0.9\textwidth]{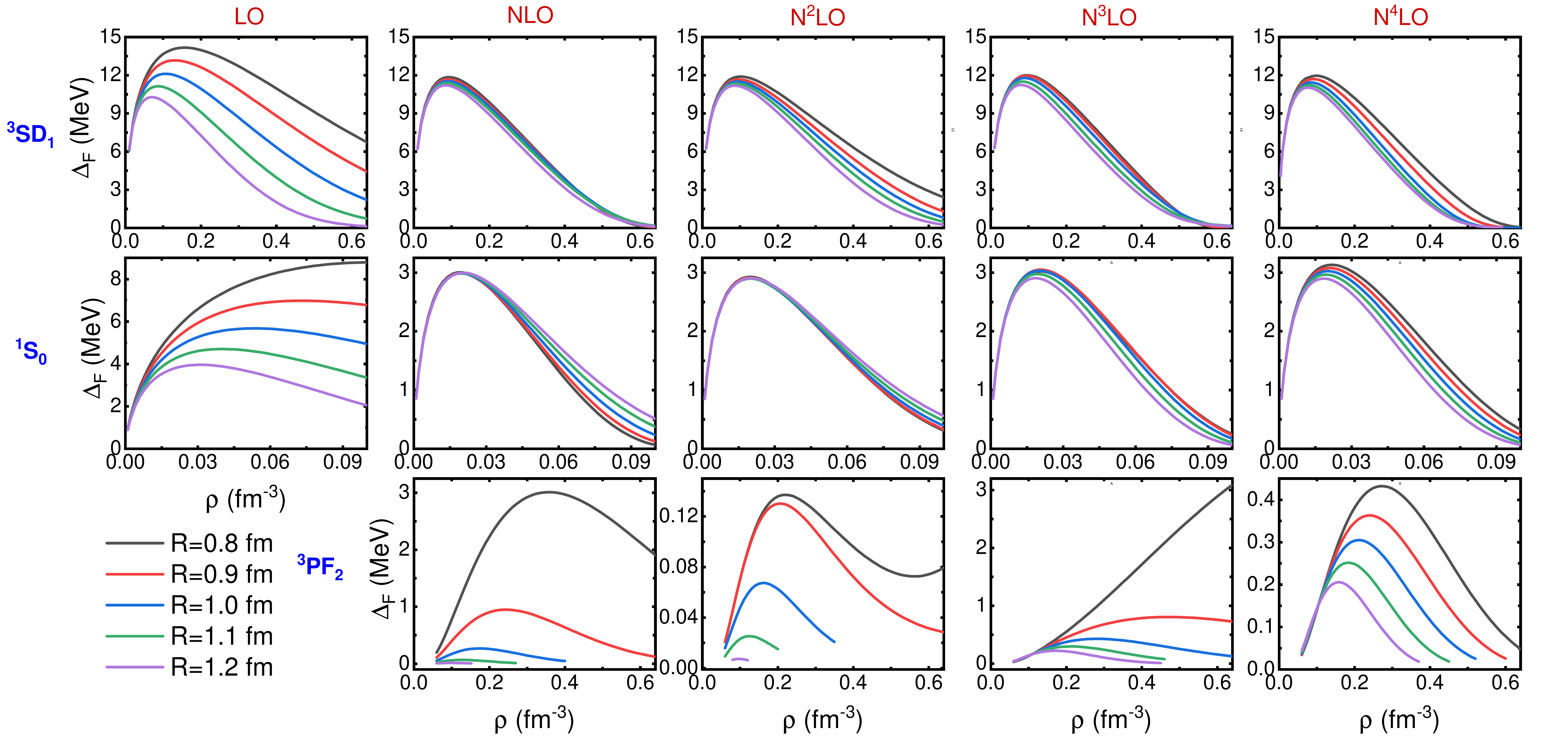}
\end{center}
\caption{(Color online) Pairing gaps in the $^3SD_1$ channel in symmetric nuclear matter (upper panels) and pairing gaps in the $^1S_0$ (middle panels) and $^3PF_2$ (lower panels) channels in neutron matter calculated with the SCS regularized chiral {\it NN} interactions from LO up through N$^4$LO for regulators $R=0.8-1.2$ fm.}\label{fig1}
\end{figure}
We present in Fig.~\ref{fig1} the pairing gaps in the isospin singlet ($T=0$) $^3SD_1$ channel (upper panels) in symmetric nuclear matter as functions of nuclear matter density $\rho$. We also show the pairing gaps in the isospin triplet ($T=1$) $^1S_0$ (middle panels) and $^3PF_2$ (lower panels) channels in pure neutron matter. We evaluate these pairing gaps in BCS approximation with the SCS regularized chiral {\it NN} interactions from LO up through N$^4$LO for regulators $R=0.8-1.2$ fm. Note that the $^1S_0$ and $^3PF_2$ pairing gaps have been calculated with the SCS regularized chiral interactions with all the regulators except for $R=0.8$ fm in Ref.~\cite{Drischler:2016cpy}. Our results are consistent with those in Ref.~\cite{Drischler:2016cpy}. We show these results for completeness.

In symmetric nuclear matter, pairing is allowed between the protons and neutrons. In upper panels of Fig.~\ref{fig1}, we observe very strong $^3SD_1$ pairing gaps of the order of $10$ MeV for all the SCS regularized chiral interactions due to the strong attraction of the {\it NN} interactions in this channel. The pairing gaps are strongly constrained by {\it NN} scattering phase shifts as investigated, e.g., in Ref.~\cite{Hebeler:2006kz}.
We find the regulator dependence of the $^3SD_1$ gaps is rather weak at low densities at each chiral order since all these interactions are able to describe {\it NN} phase shifts at low scattering energies, which correspond to low Fermi energies and equivalently low nuclear matter densities. At high densities, the regulator dependence of the $^3SD_1$ gaps becomes significant since these chiral interactions, in particular the LO interaction, are not well constrained by the {\it NN} phase shifts at high scattering energies. We notice that the pairing gaps change monotonically with the regulator for each chiral order. However, the sensitivity of the $^3SD_1$ gaps to the regulator shows no systematic trend as the chiral order increases. We observe the strongest and weakest regulator dependence of the $^3SD_1$ gaps for the LO and NLO interactions respectively, which is different from Ref.~\cite{Epelbaum:2014efa} where the regulator dependence of observables is expected to reduce going from LO to NLO/N$^2$LO and from NLO/N$^2$LO to N$^3$LO/N$^4$LO. It is noteworthy that the sensitivities of equation of states in symmetric nuclear matter and neutron matter to the regulator also show no systematic evolution with the chiral order~\cite{Hu:2016nkw}.
These complicated regulator dependence patterns may stem from different ranges of {\it NN} interactions or interplay of interactions at different ranges.

In middle panels of Fig.~\ref{fig1}, we find that the pairing gaps emerge at only low densities and the maximum pairing gaps are about $3$ MeV in the $^1S_0$ channel for all the chiral interactions except for the LO interactions. Note that we use different scales for the LO and higher-order results. The LO interactions in the $^1S_0$ channel are not able to describe {\it NN} phase shifts at even rather low scattering energies while the interactions at higher-orders are all well constrained by {\it NN} phase shifts in this channel. Therefore the LO interactions behave very differently in calculating the $^1S_0$ pairing gaps and show a strong regulator dependence, compared to the interactions at higher chiral orders. The N$^3$LO and N$^4$LO chiral interactions describe well the {\it NN} phase shifts up through scattering energy of $300$ MeV and the regulator dependence is almost invisible. However, we observe apparent regulator dependence of the $^1S_0$ gaps for the N$^3$LO and N$^4$LO interactions at even such low densities (below $0.1$ fm$^{-3}$), which could be possibly ascribed to overfitting in the presence of the redundant contact terms starting from N$^3$LO. The dependence of the $^1S_0$ gaps on the regulator increases with the density for all the chiral orders and the LO interactions show the strongest sensitivity. The sensitivity of the $^1S_0$ gaps to the regulator shows no systematic evolution with the chiral order as we observe in the $^3SD_1$ channel.

In lower panels of Fig.~\ref{fig1}, we find nonexistence of the $^3PF_2$ gaps with the SCS regularized chiral interactions at LO. Note that we use different scales for various chiral orders. The NLO and N$^2$LO interactions provide inaccurate descriptions of the {\it NN} phase shifts in the $^3PF_2$ channel from low to high scattering energies and the phase shifts show strong regulator dependence. We therefore observe apparent regulator dependence of the $^3PF_2$ pairing gaps for these two interactions from low to high densities. Since the more accurate N$^3$LO and N$^4$LO interactions with all the regulators describe well the {\it NN} phase shifts up to the scattering energy of about $200$ MeV (except for the F-wave) while their phase shifts diverge at higher energies for various regulators, the regulator dependence of the $^3PF_2$ gaps is rather weak at low densities while increases significantly with the density for these two interactions. The sensitivity of the $^3PF_2$ gaps to the regulator shows no systematic trend with the chiral order increasing as we observe in the $^3SD_1$ and $^1S_0$ channel.

\begin{figure}[tbh]
\begin{center}
\includegraphics[width=0.9\textwidth]{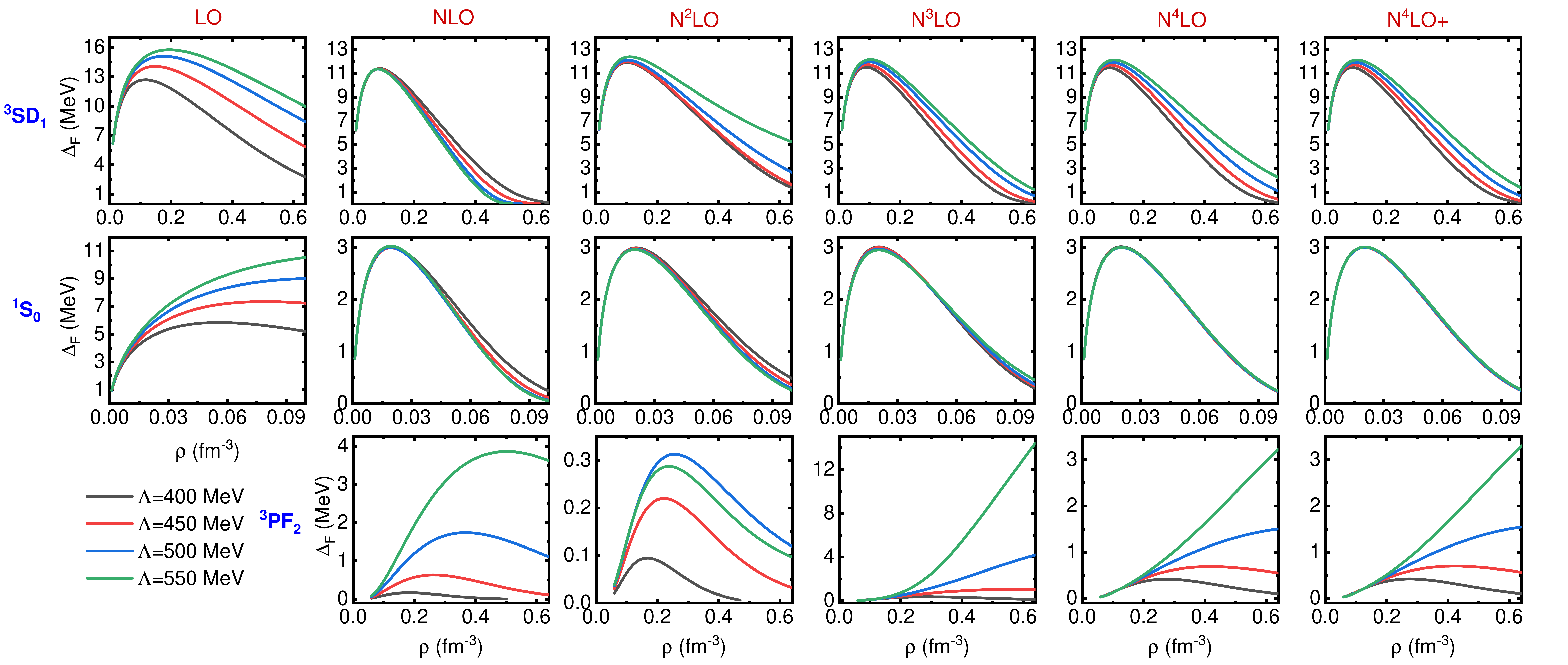}
\end{center}
\caption{(Color online) Pairing gaps in the $^3SD_1$ channel in symmetric nuclear matter (upper panels) and pairing gaps in the $^1S_0$ (middle panels) and $^3PF_2$ (lower panels) channels in neutron matter calculated with the SMS regularized chiral {\it NN} interactions from LO up through N$^4$LO+ for regulators $\Lambda=400-550$ MeV.}\label{fig2}
\end{figure}
\begin{figure}[tbh]
\begin{center}
\includegraphics[width=0.8\textwidth]{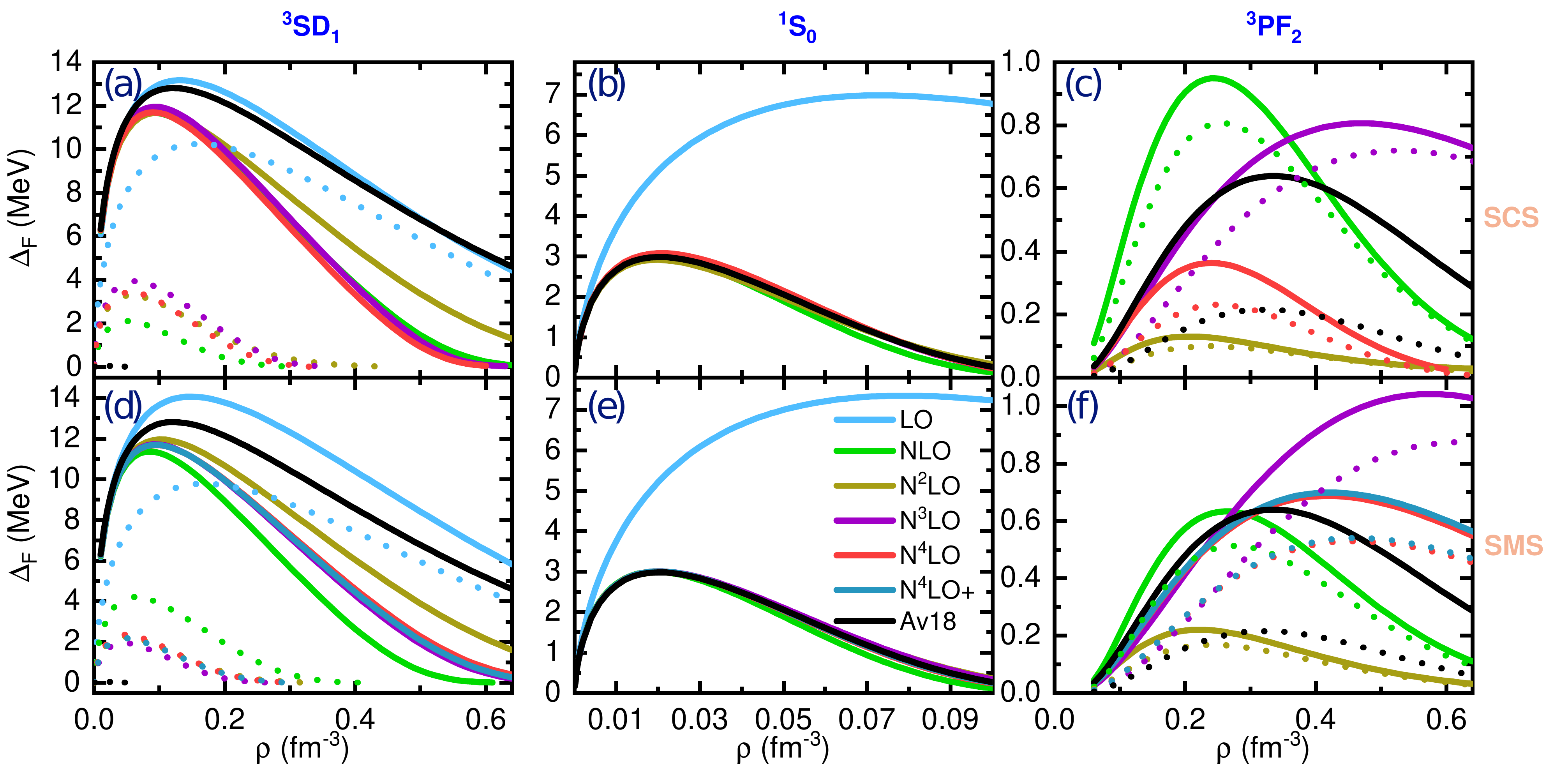}
\end{center}
\caption{(Color online) Pairing gaps (solid lines) in the $^3SD_1$ [panel (a) (d)], $^1S_0$ [panel (b) (e)], and $^3PF_2$ [panel (c) (f)] channels calculated with the Av18 potential and chiral {\it NN} interactions. Upper (lower) panels show the results of the SCS (SMS) regularized interactions from LO up through N$^4$LO (N$^4$LO+) with the same regulator $R=0.9$ fm ($\Lambda=450$ MeV). The contributions of the $^3S_1$ [panel (a) (d)] and $^3P_2$ [panel (c) (f)] single channels to the pairing gaps of the coupled $^3SD_1$ and $^3PF_2$ channels are represented by the dotted lines.}\label{fig3}
\end{figure}

Similarly, we investigate in Fig.~\ref{fig2} the pairing properties of the SMS regularized chiral interactions in $^3SD_1$ channel in symmetric nuclear matter, $^1S_0$, and $^3PF_2$ channels in neutron matter. We calculate these pairing gaps in BCS approximation from LO up through N$^4$LO+ for regulators $\Lambda=400-550$ MeV. We find in Fig.~\ref{fig2} that the density dependence and regulator dependence of the pairing gaps in the $^3SD_1$, $^1S_0$ and $^3PF_2$ channels are overall similar to those in Fig.~\ref{fig1} for the same chiral order from LO to N$^4$LO since the regulations in momentum space and in coordinate space can be approximately correlated via $\Lambda\sim\frac{1}{R}$. One of the exceptions, in contrast to the SCS case, is the sensitivity of the $^1S_0$ gaps to the regulator $\Lambda$ becomes rather weak starting from N$^3$LO and almost invisible at N$^4$LO and N$^4$LO+ due to the removal of the redundant contact terms in these SMS regularized chiral interactions. One of the significant improvements of the SMS regularized interaction, compared to the SCS regularized interactions, is including the leading F-wave contact interactions, which appear at N$^5$LO, in the N$^4$LO+ interaction. However, we find no obvious difference for the N$^4$LO and N$^4$LO+ results, even in the $^3PF_2$ channel, which will be further analyzed in Fig.~\ref{fig3}.

In Fig.~\ref{fig3}, we investigate the convergence of the pairing gaps in the $^3SD_1$, $^1S_0$, and $^3PF_2$ channels with respect to the chiral order employing the SCS and SMS regularized chiral interactions, with regulators $R=0.9$ fm and $\Lambda=450$ MeV, respectively. Each of them corresponds to one of the most accurate regularizations found in Refs.~\cite{Epelbaum:2014efa,Epelbaum:2014sza,Reinert:2017usi}. We also present the results of the Argonne v18 (Av18) potential~\cite{Wiringa:1994wb} for comparison.

We observe small difference for the $^3SD_1$ gaps of all the SCS regularized chiral interactions and the Av18 potential at low densities in Fig.~\ref{fig3} (a) since these interactions describe reasonably {\it NN} scattering phase shifts at low scattering energies. The difference become large with the density increasing since these interactions are not well constrained by the phase shifts at higher scattering energies. We notice that the $^3SD_1$ gaps tend to convergence at N$^3$LO. However, the results calculated with the most accurate N$^4$LO interaction diverge from the Av18 results at high densities, which indicates that the $^3SD_1$ gaps should be further constrained in the future.

We find in Fig.~\ref{fig3} (b) that the $^1S_0$ gaps are very close for all the SCS regularized interactions other than the LO chiral interaction since the LO interaction is not able to describe the {\it NN} phase shifts even at rather low scattering energies while all the other interactions provide good descriptions of the {\it NN} phase shifts for scattering energies up to $300$ MeV. The $^1S_0$ gaps show apparent convergence pattern with respect to the chiral order and the N$^4$LO results are very close to the Av18 results, which indicates that the $^1S_0$ gaps are well constrained by the accurate {\it NN} interactions.

In Fig.~\ref{fig3} (c) we notice that the SCS regularized chiral interactions predict different $^3PF_2$ gaps at even rather low densities. In particular, the $^3PF_2$ gap is found nonexistent for the LO interaction. We observe convergence trend for the results calculated with the chiral interactions from N$^3$LO to N$^4$LO at low densities and the converged results are consistent with the Av18 results since these three interactions describe reasonably the {\it NN} phase shifts in the $^3PF_2$ channel (regardless of F-wave) for scattering energies up to $300$ MeV. However, the convergence trend is broken at high densities, indicating that we may request higher chiral orders to reach convergence for the $^3PF_2$ pairing gaps.

In panel (d-f) of Fig.~\ref{fig3} we observe similar chiral order dependence of the $^3SD_1$, $^1S_0$, and $^3PF_2$ pairing gaps for the SMS regularized chiral interactions as in panel (a-c) for the SCS regularized chiral interactions from LO to N$^4$LO. We find in panel (d-f) that the $^3SD_1$, $^1S_0$, and $^3PF_2$ pairing gaps for the N$^4$LO and N$^4$LO+ interactions are rather close. Though the leading F-wave contact interactions of N$^5$LO level introduced in the N$^4$LO+ interaction have an small effect on the $^3PF_2$ pairing gaps, we may request a complete N$^5$LO interaction, applying to all partial waves, to evaluate the convergence pattern of the $^3PF_2$ pairing gaps.

The parameters of {\it NN} interactions adopted in this work are obtained via different fitting procedures. Therefore their detailed constituents, e.g., the off-shell constituents, could be quite different though their on-shell properties have been well confined by the same {\it NN} scattering phase shifts. These difference could be revealed in their predictions to various nuclear properties. For example, the $D-$wave probability of the deuteron calculated with these interactions are apparently different~\cite{Epelbaum:2014efa,Epelbaum:2014sza,Reinert:2017usi,Wiringa:1994wb}. In order to investigate the detailed constituents of these interactions, especially the tensor force components, we show in Fig.~\ref{fig3} the contributions of the $^3S_1$ and $^3P_2$ single channels (dotted lines) to the $^3SD_1$ and $^3PF_2$ pairing gaps.
We emphasize that the calculations with all the adopted interactions predict nonexistence of the pairing gaps in the $^3D_1$ and $^3F_2$ single channels.

As is well known, the $^3SD_1$ gap equation reduces to the Schr\"{o}dinger equation for the deuteron bound state in the limit of vanishing density~\cite{Baldo:1995zz,Lombardo:2001ek}. The accurate description of the adopted interactions of the deuteron binding energy ensures the similar behavior of the $^3SD_1$ pairing gaps at low densities in Fig.~\ref{fig3}. However, it does not mean the contributions of different components of {\it NN} interactions to the $^3SD_1$ pairing gaps are similar. Actually, the discrepancies of $^3S_1$ pairing gap among different interactions (especially the distinction between chiral interaction and Av18 potential) are remarkable, which indicates the tensor force components of these interactions in the $^3SD_1$ channel are different as expected. The difference of the tensor force effects for these interactions become more significant at higher densities.
One of the common features of the chiral interactions (regardless of the inaccurate LO interactions) and the Av18 potential is the contribution of the tensor force components are much more important than the $^3S_1$ single channel. Similarly, we find significant distinction of the tensor force effects for the adopted interactions in the $^3PF_2$ channel (see Fig.~\ref{fig3} (c) (f)). Therefore the tensor force components of these interactions in the $^3PF_2$ channel are also different. Unlike with the results in the $^3SD_1$ channel, the tensor force effects are less important than the $^3P_2$ single channel for the chiral interactions while it is opposite for the Av18 potential in the $^3PF_2$ channel.

\begin{figure}[tbh]
\begin{center}
\includegraphics[width=0.8\textwidth]{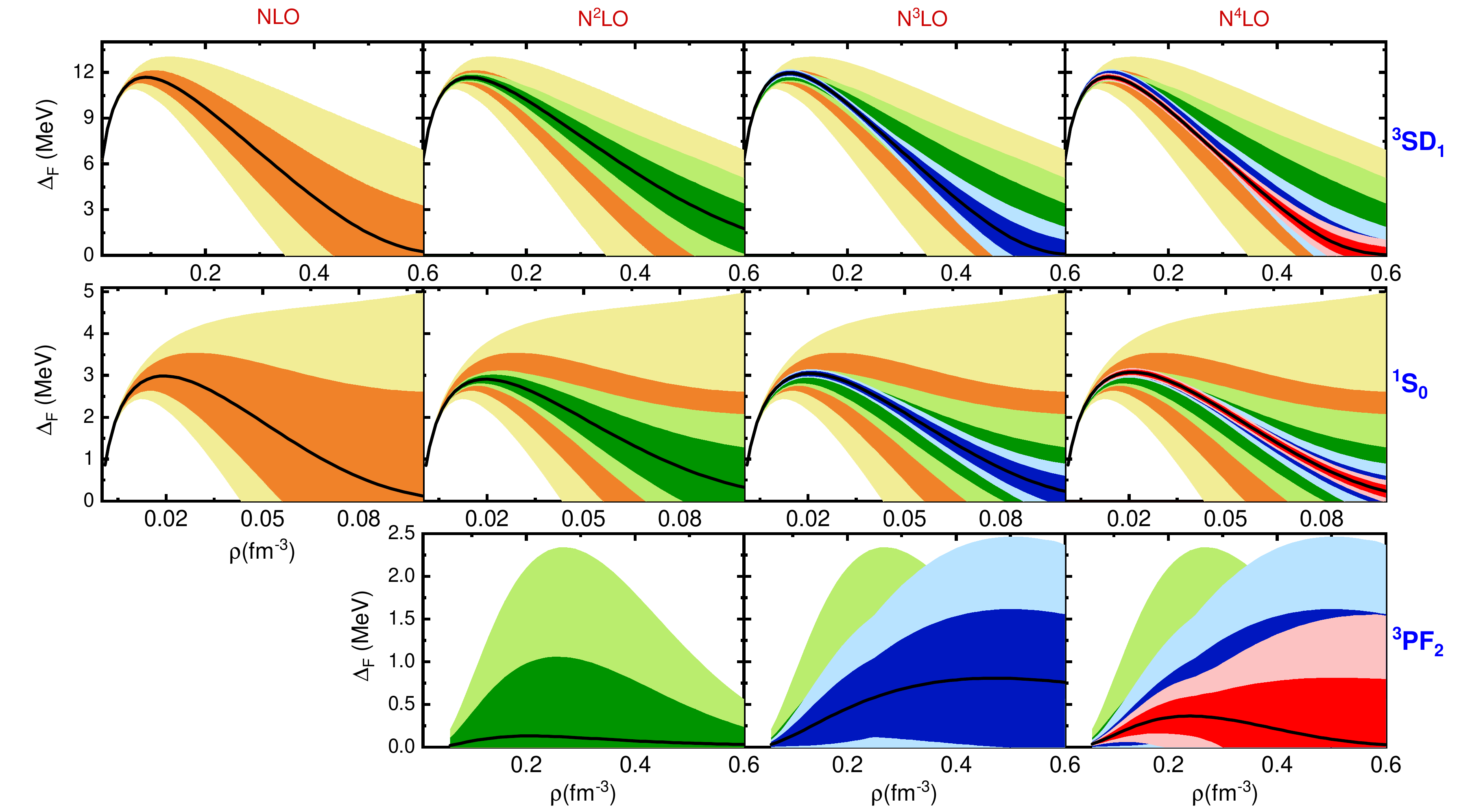}
\end{center}
\caption{(Color online) Pairing gaps with truncation errors in the $^3SD_1$, $^1S_0$, and $^3PF_2$ channels calculated by the SCS regularized chiral {\it NN} interactions with regulator $\lambda=0.9$ fm from LO up through N$^4$LO. The dark shaded band for each color indicate degree-of-belief interval is $1\sigma$, while the light ones corresponding to $2\sigma$ standard deviation.}\label{fig5}
\end{figure}
In Fig.~\ref{fig5} we estimate the truncation errors of chiral expansion for the pairing gaps calculated by the SCS regularized interactions using a Bayesian approach with the degree-of-belief intervals of $1\sigma$ and $2\sigma$ (see the appendix for details). From NLO to N$^4$LO, the truncation errors of the $^3SD_1$ and $^1S_0$ gaps decease systematically order by order. The truncation errors become rather small at N$^4$LO in particular. These calculations demonstrate that the chiral potentials in these two channels present rather good convergence for the current application. The truncation errors of the $^3PF_2$ gaps decrease also systematically order by order at low densities. However, such a systematic evolution is broken as the density increases though the truncation errors at N$^3$LO and N$^4$LO are of comparable size. It indicates that we may request higher chiral orders to reach convergence in this channel as we point out in Fig.~\ref{fig3}. The truncation errors of chiral expansion for the $^1S_0$ and $^3PF_2$ gaps calculated with the SCS regularized interactions have been investigated in Ref.~\cite{Drischler:2016cpy} with an easily operational analysis methodology proposed in Refs.~\cite{Epelbaum:2014efa,Epelbaum:2014sza}. These evaluations neglect the LO contributions to the higher-order uncertainties and a term ensuring that the next order always lies within the uncertainty band of the previous order in contrast to Refs.~\cite{Epelbaum:2014efa,Epelbaum:2014sza}. Therefore the systematic evolution of the truncation errors for the $^1S_0$ gaps with the chiral order we observe in Fig.~\ref{fig3} was not found in Ref.~\cite{Drischler:2016cpy}. The systematic evolution of the truncation errors for the $^3PF_2$ gaps with the chiral order at low densities we observe in Fig.~\ref{fig3} was also not found in Ref.~\cite{Drischler:2016cpy}. We are consistent with Ref.~\cite{Drischler:2016cpy} that the uncertainties are very small for the $^1S_0$ channel but sizable for the $^3PF_2$ channel. We find similar behavior for the truncation errors obtained with the SMS regularized interactions (see the appendix for details). Since the N$^4$LO+ interaction is not a complete N$^5$LO interaction, we do not evaluate the truncation errors of pairing gaps at N$^4$LO+.

We emphasize that we investigate the pairing properties of the two-nucleon forces and do not include the contributions of three-nucleon forces in this work. The pairing gaps and the truncation errors starting from N$^2$LO are incomplete and should be revisited once the calculations with the three-nucleon forces become available. The results at N$^2$LO and beyond obtained in this work may reveal a potentially achievable accuracy at the corresponding chiral orders.

\section{conclusions and outlook}

In conclusion, we investigated the pairing properties of state-of-the-art SCS and SMS regularized chiral EFT interactions in nuclear matter within the BCS approximation. Specifically, we calculated the pairing gaps in the $^3SD_1$, $^1S_0$, and $^3PF_2$ channels.

We investigated the regulator dependence of the pairing gaps for the SCS regularized chiral interactions. The $^3SD_1$ and $^1S_0$ pairing gaps show weak regulator dependence at low densities but reveal apparent regulator dependence as the density increases. We found similar behavior for the $^3PF_2$ pairing gaps at N$^3$LO and N$^4$LO while the NLO and N$^2$LO results show an overall strong regulator dependence from low to high densities. We found roughly similar regulator dependence for the results of the SMS regularized chiral interactions.  One of the exceptions, in contrast to the SCS case, is that the sensitivity of the $^1S_0$ gaps to the regulator  becomes rather weak starting from N$^3$LO and almost invisible at N$^4$LO and N$^4$LO+ due to the removal of the redundant contact terms in these SMS regularized chiral interactions.

We further investigated the convergence of the pairing gaps of the chiral interactions with respect to the chiral order. The $^3SD_1$ and $^1S_0$ pairing gaps are overall converged from low to high densities while the $^3PF_2$ results are converged at only low densities. The converged results of the chiral interactions at low densities coincide with the Av18 results for these three channels. However, we observed apparent discrepancy for the chiral interaction and Av18 potential in the $^3SD_1$ and $^3PF_2$ channels at high densities, indicating the pairing gaps in these two channels should be further constrained in the future. We found similar chiral order dependence for the SMS regularized chiral interactions. The leading F-wave contact interactions of N$^5$LO level introduced in N$^4$LO+ interaction are insufficient to provide complete convergence for the $^3PF_2$ pairings.

In addition, we have investigated the effect of the tensor force on the $^3SD_1$ and $^3PF_2$ pairing gaps with the Av18 potential and the chiral interactions. We found different tensor force effects for the $^3SD_1$ and $^3PF_2$ pairing gaps and such divergence becomes more significant as the density increases. We therefore concluded that the tensor force components in these interactions are quite different. One of the common features of the chiral interactions (regardless of the inaccurate LO interactions) and the Av18 potential is the contribution of the tensor force components are overall more important than the $^3S_1$ single channel. In contrast to the $^3SD_1$ channel, the tensor force effects are less important than the $^3P_2$ single channel for the chiral interactions while it is opposite for the Av18 potential in the $^3PF_2$ channel.

Finally, we estimated the truncation errors of chiral expansion of the pairing gaps using a Bayesian approach. We found systematic reduction of the truncation errors from NLO to N$^4$LO for the $^3SD_1$ and $^1S_0$ pairing gaps, indicating the chiral interactions in these two channels show rather good convergence. The truncation errors of the $^3PF_2$ gaps reduce also systematically order by order at low densities. However, such a systematic evolution is broken as the density increases though the truncation errors at N$^3$LO and N$^4$LO are of comparable size, which supports our conclusion that we may request higher chiral orders in this channel.

In this work, we used free single particle spectrum which would be corrected by the nucleon effective mass, depletion of the Fermi surface due to short-range correlations and the medium polarization effects in more realistic nuclear matter. We will take these corrections into account with the many-body Brueckner Hartree Fock (BHF) theory in the future. We adopted only two-body nuclear force (2BF) in the current calculations. The expressions for the three-body force (3BF) have been worked out completely up to N$^3$LO. We will include the chiral 3BF in the BHF theory and investigate the effects of the 3BF on the pairing correlations in nuclear matter, which is challenging in numerical implementations. Employing self-consistent 2BF and 3BF, we will be able to study the effect of the pairing correlations in the neutron star cores on the neutron star cooling phenomena.

\section*{Acknowledgments}
This work were supported by the National Natural Science Foundation of China (Grant Nos. 11975282, 11705240, 11435014), the Strategic Priority Research Program of Chinese Academy of Sciences, Grant No. XDB34000000, the Key Research Program of the Chinese Academy of Sciences under Grant No. XDPB15, DFG and NSFC through funds provided
to the Sino-German CRC 110 ``Symmetries and the Emergence of Structure in QCD'' (NSFC Grant No. 12070131001, Project ID 196253076-TRR 110), and ERC Nuclear Theory (Grant No. 885150).

\appendix
\renewcommand{\appendixname}{Appendix}

\section{Bayesian analysis}
\label{sec:appendixB}
We use the Bayesian scheme of Refs.~\cite{Furnstahl:2015rha,Melendez:2017phj} to estimate the truncation errors of pairing gaps from chiral potentials. The generic assumption is a nuclear observable $X$ in Chiral EFT can be expanded with a dimensionless parameter $Q$ as follows:
\begin{eqnarray}
 X&=&X_{ref}\sum_{n=0}^{\infty} c_n Q^n,
\label{eq:observe}
\end{eqnarray}
where $X_{ref}$ is the natural size of $X$ and $c_n$s are dimensionless parameters.
In this work, we investigate the truncation errors of the pairing gap $\Delta_F$ in nuclear matter. Therefore the observable $X$ is $\Delta_F$ and the expansion parameter is regarded as $Q=\frac{k_F}{\Lambda_b}$, with $k_F$ the Fermi momentum of nucleon determined by the nuclear density $\rho$ and $\Lambda_b$ the Chiral EFT breakdown scale. We take $\Lambda_b=700$ MeV, which is much higher than the maximum Fermi momentum $515$ MeV (corresponding to $\rho=0.6$ fm$^{-3}$ for pure neutron matter) in this work.

The error of the observable truncated at the order $k$ of the expansion is defined as $X_{ref}\Delta_k$, with the dimensionless function $\Delta_k$ calculated by
\begin{eqnarray}
 \Delta_k&=&\sum_{n=k+1}^{\infty} c_n Q^n.
\label{eq:deltak}
\end{eqnarray}
In practice, we sum over $n$ up to $h+k+1$ order and neglect the higher orders.
The coefficients $c_n$ with $n\ge k+1$ are extracted by the known expansion coefficients $c_n$ with $n\le k$. In Bayesian model, we define a probability distribution function (pdf) for $\Delta_k$ as $pr_h(\Delta|\bm{c_k})$, determined by a vector composed of lower-coefficients, $\bm{c_k}\in \{c_2, c_3,\cdots, c_k \}$. The subscript $h$ means only $h$ higher-terms are included in the truncation error, which is $10$ in this work. Note that $\bm{c_k}$ does not include $c_0$ and $c_1$ since $c_0$ is dependent on the natural size of $X$ and $c_1=0$ required by the symmetry in Chiral EFT.

The pdf determines the degree-of-belief (DoB), $p$, with the highest posterior density (HPD),
\begin{eqnarray}
 p&=&\int_{-d_k^{(p)}}^{d_k^{(p)}} pr_h(\Delta|\bm{c_k})d\Delta,
\label{eq:dob}
\end{eqnarray}
where $(100\times p)\%$ is the probability for the true value of the nuclear observable $X$ staying in $\pm X_{ref}d_k^{(p)}$ at the $(k+1)$ order (N$^k$LO) prediction.

In Ref.~\cite{Furnstahl:2015rha}, $\Delta_k$ was derived in terms of the expansion coefficients $c_n$s by assuming them as random variables drawn from a shared distribution centered at zero with a characteristic size or upper bound $\bar c$.
The pdf function can be written with Bayesian theorem as
\begin{eqnarray}
 pr_h(\Delta|\bm{c_k})&=&\frac{\int_{0}^{\infty} d\bar c pr_h(\Delta|\bar c)pr(\bar c) \prod_{n=2}^k pr(c_n|\bar c)}{\int_{0}^{\infty} d\bar c pr(\bar c) \prod_{n=2}^k pr(c_n|\bar c)},
\label{eq:prh}
\end{eqnarray}
where we use the following priors
\begin{eqnarray}
 pr(c_n|\bar c)&=&\frac{1}{2\bar c}\theta(\bar c-|c_n|), \\ \nonumber
pr(\bar c)&=&\frac{1}{\sqrt{2\pi}\bar c\sigma}e^{-(\ln\bar c)^2/2\delta^2}.
\label{eq:prc}
\end{eqnarray}
The prior $pr_h(\Delta|\bar c)$ can be worked out with

\begin{eqnarray}
 pr_h(\Delta|\bar c)&=&\frac{1}{2\pi} \int_{-\infty}^{\infty} dt \cos (\Delta t) \prod_{i=k+1}^{k+h}\frac{\sin (\bar c Q^i t)}{\bar c Q^i t}.
\label{eq:prdeltac}
\end{eqnarray}
With the above equations, we can obtain $d_k^{(p)}$ in Eq.~\ref{eq:dob} numerically as an inversion problem.

In this work, we take $X_{ref}$ to be $\Delta_F$ of the LO interactions for the $^3SD_1$ and $^1S_0$ channels. Since we find nonexistence of $^3PF_2$ pairing gaps for the LO interactions, we take $X_{ref}$ to be $\Delta_F/Q^2$ of the NLO interactions in this channel.

\section{Truncation errors of pairing gaps with the SMS regularized interactions}
\label{sec:appendixC}

\begin{figure}[tbh]
\begin{center}
\includegraphics[width=0.8\textwidth]{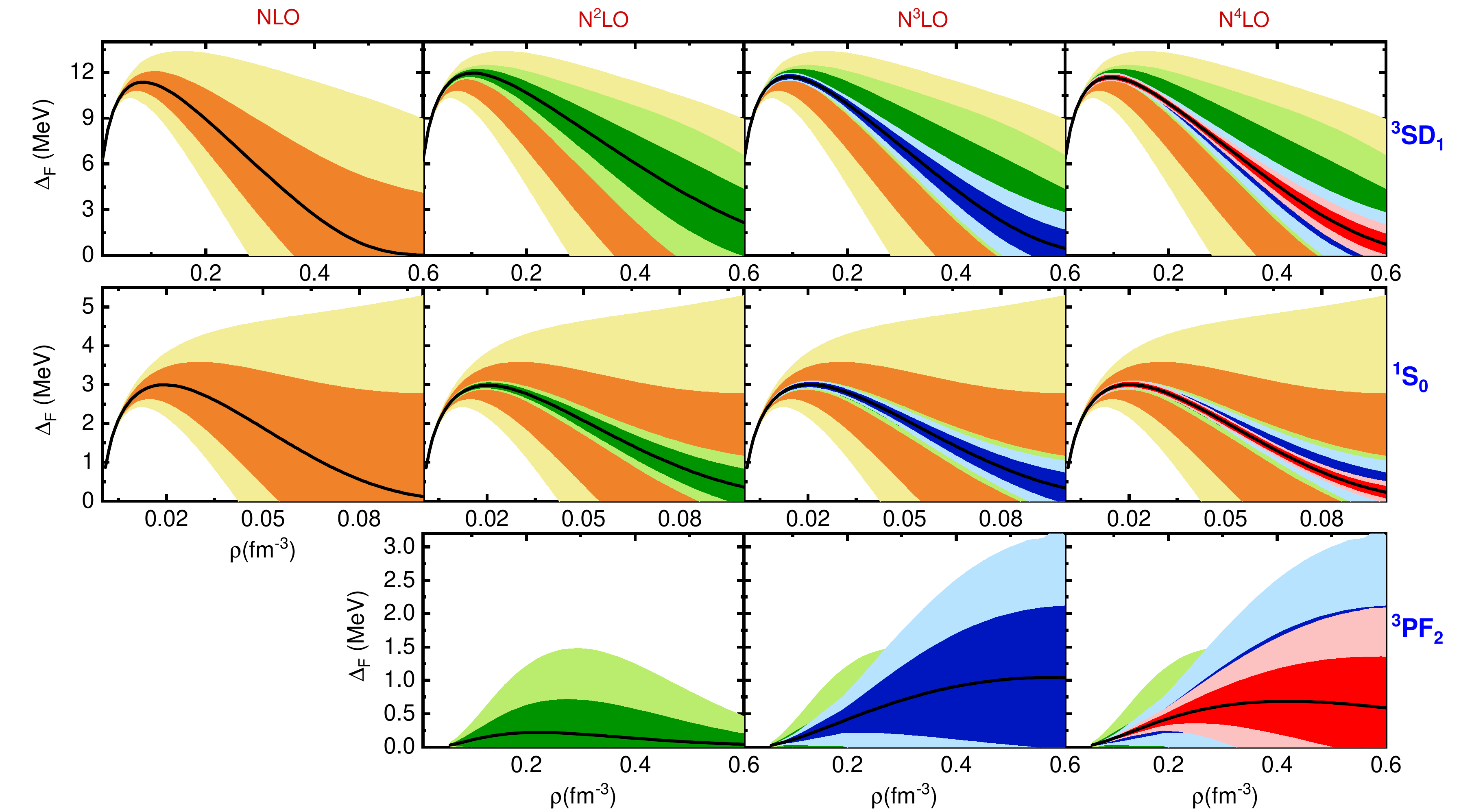}
\end{center}
\caption{(Color online) Pairing gaps with truncation errors in the $^3SD_1$, $^1S_0$, and $^3PF_2$ channels calculated by the SMS regularized chiral {\it NN} interactions with regulator $\Lambda=450$ MeV from LO up through N$^4$LO. The dark shaded band for each color indicate degree-of-belief interval is $1\sigma$, while the light ones corresponding to $2\sigma$ standard deviation.}\label{fig6}
\end{figure}

\end{document}